\documentclass{autart_mod}

\usepackage{natbib} 
\usepackage{amsmath} 
\usepackage{amssymb} \usepackage{bm}
\usepackage{epsfig} \usepackage{graphicx} \usepackage{float}

\newcommand{\ou}[2]{\raisebox{0.35ex}{$\genfrac{}{}{0pt}{3}{#1}{#2}$}}
\def\cal{\fam2 }
\begin{document}


\begin{frontmatter}
  
  \title{Cosmic rays at ultra high energies (Neutrinos!)}
  
  \author{Markus~Ahlers}, \author{Andreas~Ringwald} \address{Deutsches
    Elektronen-Synchrotron DESY, Notkestr. 85, D-22607 Hamburg, Germany}
  
  \author{Huitzu~Tu} \address{University of Southern Denmark, Campusvej 55,
    DK-5230 Odense M, Denmark}

\begin{abstract}
  Resonant photopion production with the cosmic microwave background predicts
  a suppression of extragalactic protons above the famous
  Greisen--Zatsepin--Kuzmin cutoff at about $E_\mathrm{GZK} \approx
  5\times10^{10}$~GeV.  Current cosmic ray data measured by the AGASA and
  HiRes Collaborations do not unambiguously confirm the GZK cutoff and leave a
  window for speculations about the origin and chemical composition of the
  highest energy cosmic rays.  In this work we analyze the possibility of
  strongly interacting neutrino primaries and derive model-independent
  quantitative requirements on the neutrino-nucleon inelastic cross section
  for a viable explanation of the cosmic ray data.  Search results on weakly
  interacting cosmic particles from the AGASA and RICE experiments are taken
  into account simultaneously.  Using a flexible parameterization of the
  inelastic neutrino-nucleon cross section we find that a combined fit of the
  data does not favor the Standard Model neutrino-nucleon inelastic cross
  section, but requires, at 90\% confidence level, a steep increase within one
  energy decade around $E_\mathrm{GZK}$ by four orders of magnitude.  We
  illustrate such an enhancement within some extensions of the Standard Model.
  The impact of new cosmic ray data or cosmic neutrino search results on this
  scenario, notably from the Pierre Auger Observatory soon, can be immediately
  evaluated within our approach.
\end{abstract}

\end{frontmatter}

\section{Introduction}

In the mid-1960's Greisen~\cite{Greisen:1966jv}, Zatsepin and
Kuzmin~\cite{Zatsepin:1966jv} realized that the space-filling molasses of
photons constituting the cosmic microwave background (CMB) limits the
observation of high-energy charged particles originating at astrophysical
sources.  The attenuation length of protons within the CMB drops below 50 Mpc
above the photo-pion production threshold of about $E_\mathrm{GZK}\approx
5\times10^{10}$~GeV. In the case of heavier nuclei, photo-disintegration with
CMB photons predicts a similar and even stronger attenuation above this energy
(e.g.,~\cite{Nagano:2000ve}). Hence, the apparent horizon of charged ultra
high energy (UHE) cosmic rays (CRs) is of size comparable with the diameter of
our local supercluster.  Accordingly, the contribution of extragalactic
charged particles to the CR spectrum measured on Earth is expected to show a
cutoff at $E_\mathrm{GZK}$.

\begin{figure}[t]
\begin{center}
  \includegraphics[width=\linewidth,clip=true]{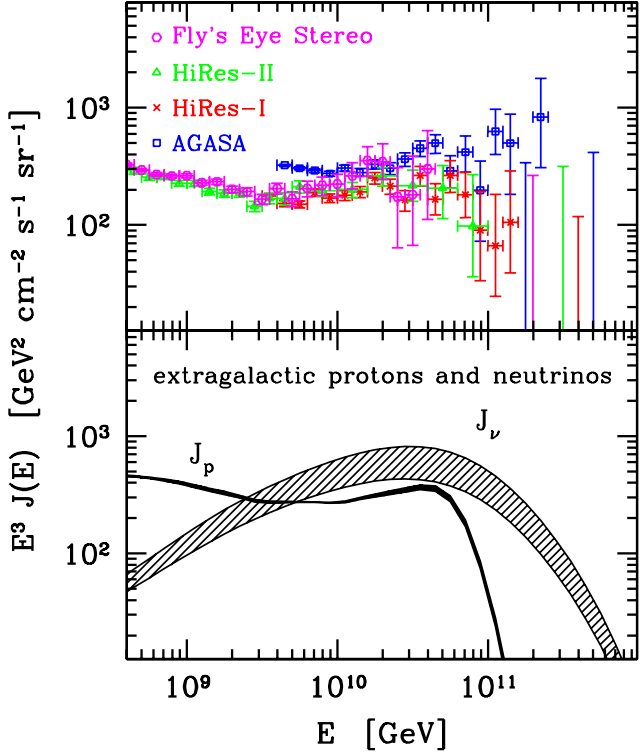}
\end{center}
\caption{\label{data}{\it Top panel:} The quasi-vertical flux of UHE CRs measured by AGASA, Fly's Eye and HiRes-I/II. {\it Bottom panel:} The shaded band shows the flux of
  extragalactic protons corresponding to the 90\% CL of our fit. For the
  normalization we averaged the data between $10^{8.6}\,\mathrm{GeV}$ and
  $10^{11}\,\mathrm{GeV}$ according to the experimental exposure (see \S
  \ref{CRS}).  The hatched band shows the combined flux of cosmogenic
  neutrinos and neutrinos from optically thin sources.}
\end{figure} 

The UHE CR spectrum measured by the AGASA~\cite{Takeda:2002at} and
HiRes~\cite{Bird:1994wp,Abbasi:2005ni,Abbasi:2005bw,HURL} Collaborations is
shown in Fig.~\ref{data}.  The reported flux estimates derived from UHE events
show a qualitative difference between the data sets.  While AGASA reports an
excess of eleven events above $10^{11}$~GeV, the HiRes data seem to support
the existence of the GZK-cutoff.  However, the significance of the difference
is small due to the low statistics as well as large systematic errors in
energy calibration. Apart from this, the origin and chemical composition of
UHE CR remains a puzzle in astroparticle physics.  The Pierre Auger
Observatory (PAO), which will publish its first results in summer 2005, is
expected to shed light on these questions.

A recent analysis of the HiRes data~\cite{Bergman:2004bk} indicates a change
in the UHE CR composition at around $10^{8.6}$~GeV from heavy nuclei to a
light component, which can be interpreted as the onset of extragalactic proton
dominance in the data. If this is correct the UHE CR data is expected to show
a cutoff at $E_\mathrm{GZK}$ due to resonant photo-pion production in the CMB.
This is shown as the thin shaded band in Fig.~\ref{data} for our model of
extragalactic protons described in \S \ref{CRS}.  On the other hand, this
mechanism leads to the generation of high energy neutrinos from the decay of
charged pions. This flux of {\it cosmogenic
  neutrinos}~\cite{Beresinsky:1969qj,Stecker:1978ah} may exceed the proton
flux above and around $E_\mathrm{GZK}$ depending on the particular model
parameters.  In general, these cosmogenic neutrinos are accompanied by
neutrinos originating from photo-hadronic processes directly in the CR
sources. Recently, two of us have been engaged in a derivation of a lower
bound on this additional source of neutrinos directly related to the observed
flux of CRs~\cite{Ahlers:2005sn}.  The sum of these two contributions is shown
as the hatched band in Fig.~\ref{data}.

The observation that the UHE cosmic neutrino flux may exceed the proton flux
above $E_{\rm GZK}$ has motivated Berezinsky and
Zatsepin~\cite{Beresinsky:1969qj} to speculate on the possibility that
neutrinos constitute the highest energy CR events, assuming them to
interact so strongly as hadrons at the relevant energies~\cite{note3}.  Quantitatively,
this requires a (rapid) rise of the neutrino-nucleon total inelastic cross
section $\sigma_{\nu N}^{\rm in}$ by at least five orders of magnitude, above the Standard Model (SM)
prediction, $\sigma^\mathrm{SM}_{\nu N}\approx 7.84\,\mathrm{pb}\,
\left(E_\nu/\mathrm{GeV}\right)^{0.363}$~\cite{Gandhi:1998ri,Kwiecinski:1998yf}.
The realization of such a behavior has been proposed abundantly in scenarios
beyond the (perturbative) SM: e.g. arising through
compositeness~\cite{Domokos:1986qy,Bordes:1997bt,Bordes:1997rx}, through
electroweak
sphalerons~\cite{Aoyama:1986ej,Ringwald:1989ee,Espinosa:1989qn,Khoze:1991mx,Ringwald:2002sw,Bezrukov:2003qm,Ringwald:2003ns,Fodor:2003bn,Han:2003ru},
through string excitations in theories with a low string and unification
scale~\cite{Domokos:2000dp,Domokos:2000hm,Burgett:2004ac}, through
Kaluza-Klein modes from compactified extra
dimensions~\cite{Domokos:1998ry,Nussinov:1998jt,Jain:2000pu,Kachelriess:2000cb,Anchordoqui:2000uh,Kisselev:2003rz},
or through $p$-brane production in models with warped extra
dimensions~\cite{Ahn:2002mj,Jain:2002kf,Anchordoqui:2002it}, respectively (for
a recent review, see Ref.~\cite{Fodor:2004tr}).

Note that neutrino-nucleon inelastic cross sections are in general constrained
by the search results on normal weakly-interacting UHE neutrinos.  Up to now,
UHE cosmic neutrinos have been searched for in the Earth atmosphere (Fly's
Eye/AGASA), in the Greenland (FORTE~\cite{FORTE}) and Antarctic ice sheet
(AMANDA~\cite{AMANDA}/RICE~\cite{RICE}), in the sea/lake
(BAIKAL~\cite{BAIKAL}), or in the regolith of the moon (GLUE~\cite{GLUE}).
For a given flux, the low if not zero search results so far can be turned into
model-independent upper bounds on the neutrino-nucleon inelastic cross section
in the energy range where $\sigma^\mathrm{SM}_{\nu N}\lesssim\sigma_{\nu
  N}^{\rm in}\lesssim 0.05-0.5$~mb~\cite{Anchordoqui:2004ma}. In strongly-interacting
neutrino scenarios, this demands that the neutrino-nucleon inelastic cross
section should pass very rapidly through this intermediate energy range.
This constraint will be further strengthened with PAO's search results on
deeply-penetrating showers (see e.g.~\cite{Anchordoqui:2004ma})
, complemented by IceCube~\cite{ICECUBE}, ANITA~\cite{ANITA} and possibly by
EUSO~\cite{EUSO}, SalSA~\cite{SALSA}, and OWL~\cite{OWL}.

In this work we present a general method, which was previously exploited in
Ref.~\cite{Fodor:2003bn}, to derive model-independent quantitative
requirements on the cross section in strongly interacting neutrino scenarios.
It exploits current CR data and UHE neutrino search results, and can
easily incorporate or be applied to other data-releasing experiments, notably
PAO. With this method the central question of the viability of strongly
interacting neutrino scenarios can be addressed, confining
the requirements on astrophysics (CR sources) and particle physics (inelastic
$\nu N$ cross section) simultaneously.

The outline of the paper is as follows. In \S \ref{CRS} we calculate the
incident fluxes of extragalactic protons and neutrinos assuming a power-like
proton injection spectrum and a cosmic evolution of the source luminosity.
The attenuation of the initial fluxes due to interactions with CMB photons and
adiabatic energy losses are taken into account with the help of propagation
functions.  The number of neutrino-induced shower and cascade events measured
at different zenith angles will depend strongly on the probability for
neutrino scattering on target nucleons. The effect of a varying
neutrino-nucleon inelastic cross section on the measurement by different type
of experiments will be discussed in \S \ref{nueas}.  In \S \ref{UHECS} we
adopt a flexible parameterization for the rise of the inelastic
neutrino-nucleon cross section with energy.  This allows us to derive
quantitative criteria on the viability of the strongly-interacting neutrino
scenarios.  In \S \ref{conclusion} we present our conclusions and give an
outlook.

\section{\label{CRS}Extragalactic Particle Production}

\begin{figure}[t]
\begin{center}
  \includegraphics[width=0.49\linewidth,clip=true]{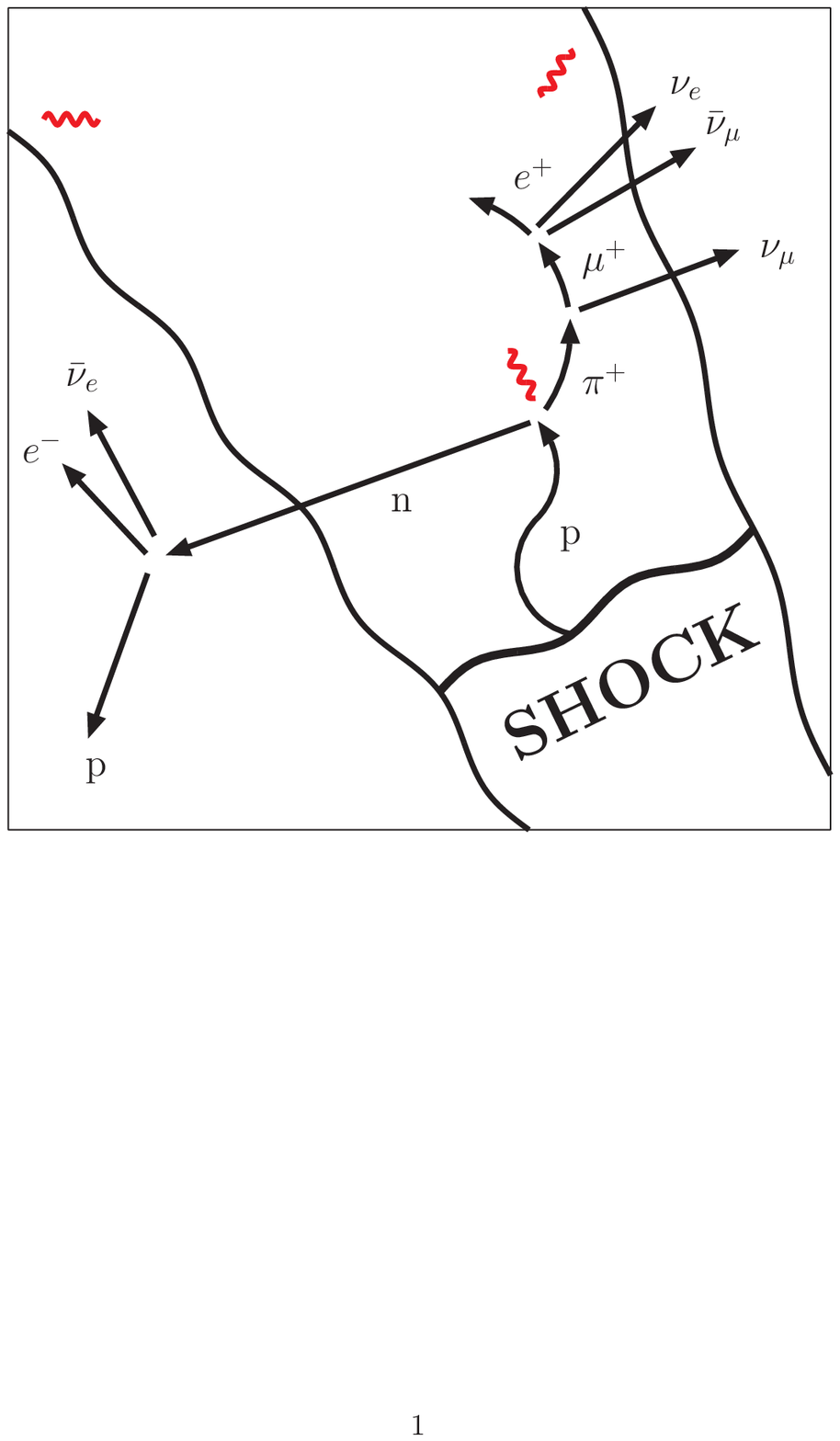} \hfill
  \includegraphics[width=0.49\linewidth,clip=true]{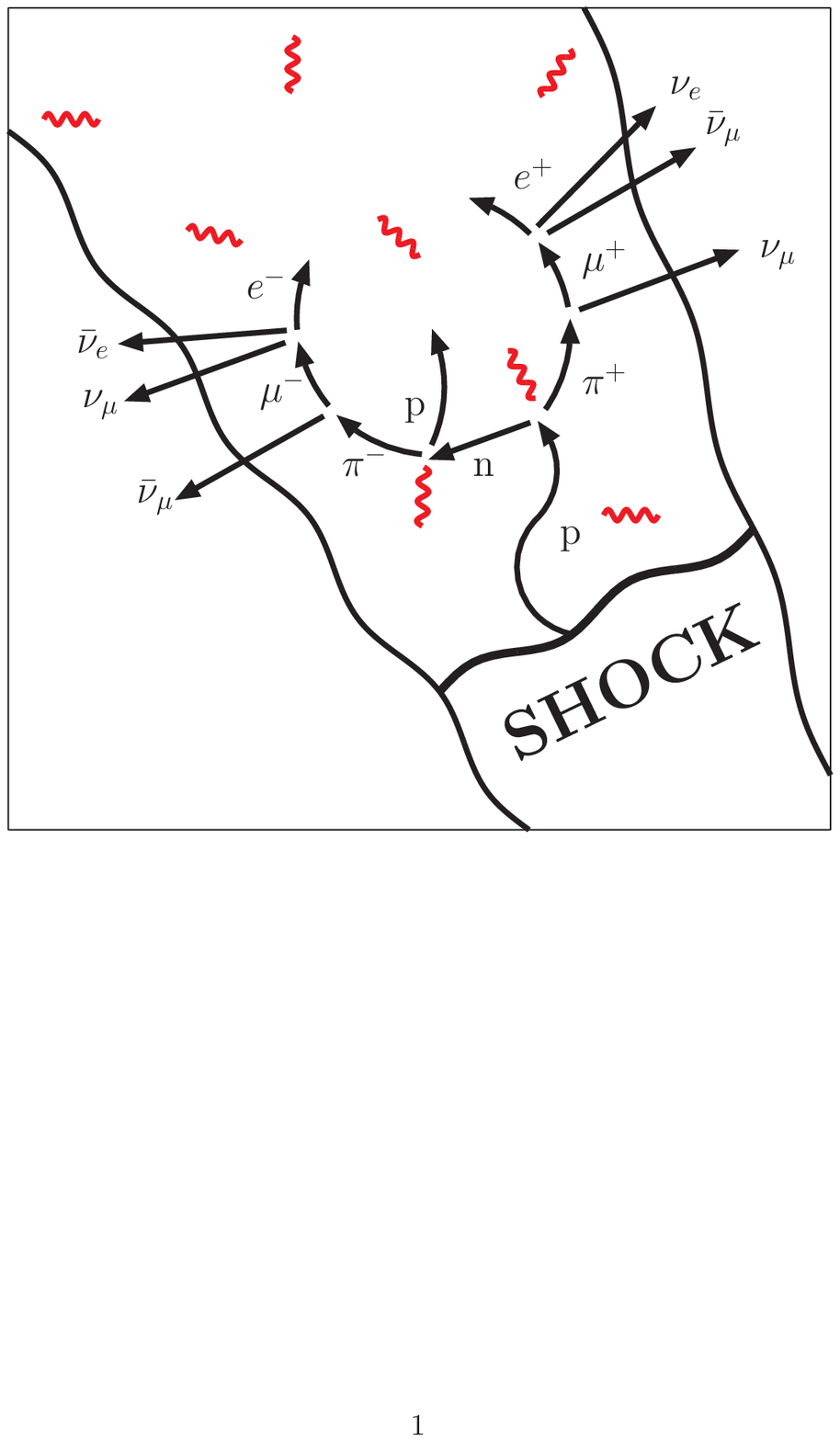}
\end{center}
\caption{\label{thin}Two possible production mechanisms of CRs by beam protons
  in a cosmic accelerator. The relative fluxes depend on the optical thickness
  of the magnetic confinement region.}
\end{figure}

Following typical models of cosmic particle accelerators, such as active
galactic nuclei or gamma ray bursts (see e.g.~\cite{Torres:2004hk} for a nice
review), we assume that the origin of extragalactic protons are
$\beta$-decayed neutrons, which are produced by photo-hadronic processes of
beam protons and subsequently escape from the magnetically confined
acceleration region.  Fig.~\ref{thin} shows a sketch of this mechanism.  The
relative output of other neutral particles, in particular high energy
neutrinos, depend on the details of the source such as the densities of the
target photons and the ambient gas~\cite{Mannheim:1998wp}.  In the following
we will not be concerned about the detailed accelerator specifications and
start directly with the injection spectrum of neutrons.

\subsection{\label{pflux}Extragalactic Protons}
The apparent isotropy of the CR data suggests the assumption of a spatially
homogeneous and isotropic distribution of the extragalactic neutron sources.
Furthermore we assume that the emissivity distribution per co-moving volume
$\mathcal{L}_n$ factorizes into the red-shift evolution of the source
luminosity and the injection spectrum.  For simplicity we choose a widely used
power-law ansatz with an exponential cut-off (see, e.g.
Refs.~\cite{Protheroe:1995ft,Engel:2001hd,Fodor:2003ph,Semikoz:2003wv,Berezinsky:2004fk}):
\begin{gather}
  \mathcal{L}_n(z,E_n) \propto
  (1+z)^n\,E_n^{-\gamma}\,e^{-E_n/E_\mathrm{max}}\\
  \nonumber z_\mathrm{min}<z<z_\mathrm{max}
\end{gather}
We exclude nearby (redshift $z_\mathrm{min}$) and early ($z_\mathrm{max}$)
sources and fix these parameters in the following at $z_\mathrm{min} = 0.012$
(corresponding to $r_\mathrm{min} \approx 50\, \mathrm{Mpc}$) and
$z_\mathrm{max} = 2$.  The maximal injection energy $E_\mathrm{max}$ is fixed
at $10^{12}$~GeV in our analysis.

The relevant flux of protons propagating towards the Earth resulting from the
$\beta$-decay of the source neutrons is attenuated due to energy red-shift,
$e^+e^-$ pair production and photo-pion production with CMB photons (see
e.g.~\cite{Engel:2001hd,Fodor:2003ph,Semikoz:2003wv}). The last of these
processes provides the source of cosmogenic neutrinos. If we define
$P_{p|n}(E;E_n,r)$ as the expected number of protons above an energy $E$ given
a neutron with injection energy $E_n$ at a distance $r$, then the expected
proton flux incident on Earth can be expressed as
\begin{equation}
\label{jp_n}
  J_p(E) = \frac{1}{4\pi}\int\limits_0^\infty \d
  E_n\int\limits_0^{\infty} \d r
  \left|\frac{\partial P_{p|n}(E;E_n,r)}{\partial
  E}\right|\mathcal{L}_n\, .
\end{equation}
The propagation functions $P_{a|b}$~\cite{Fodor:2000yi,Fodor:2003bn} have been
calculated using the SOPHIA Monte-Carlo program~\cite{Mucke:1999yb}, and are
available at
{\tt www.desy.de/\symbol{126}uhecr/}.  The propagation distance $r$ and the
redshift $z$ are related by $\d z = (1+z)H(z)\d r$, where the Hubble expansion
rate at a redshift $z$ is related to the present one $H_0$ through $H^2 (z) =
H^2_0\, \left[\Omega_M (1 + z)^3 + \Omega_\Lambda \right]$.  Following recent
results in cosmology \cite{Tegmark:2003ud} we will assume a flat
$\Lambda$-dominated universe with relative energy densities
$\Omega_\mathrm{M}=0.3$ and $\Omega_\Lambda=0.7$ as well as $H_0 =
72\,\mathrm{km}\,\mathrm{s}^{-1}\,\mathrm{Mpc}^{-1}$, the center value of the
Hubble Space Telescope Key project~\cite{Freedman:2000cf}.

\subsection{\label{nuflux}Cosmogenic Neutrinos and Neutrinos from Optically Thin Sources}

Isospin and charge conservation requires that each neutron produced in
photohadronic interactions of beam protons is accompanied by a charged pion.
It decays and produces $\nu_\mu, \bar{\nu}_\mu$ and $\nu_e$'s, as sketched in
Fig.~\ref{thin}. On average, each neutrino carries about a quarter of the
pion's energy. If we define $\epsilon_\pi$ as the ratio $E_{\pi^+}/E_n$
(see~Ref.~\cite{Waxman:1998yy}) we can express the average energy of a single
neutrino as $E_\nu \approx \epsilon_\pi E_n/4$.  In the following we will use
$\epsilon_\pi\approx0.28$, suitable for resonant photoproduction at the
energies in question (see Ref.~\cite{Ahlers:2005sn}).

In optically thin sources the relative normalization of the neutron and
neutrino emissivity is fixed by relating their bolometric fluxes per co-moving
volume and is given by
\begin{equation}
  \frac{\epsilon_\pi}{4}\mathcal{L}_\nu(z,\frac{\epsilon_\pi}{4} E_n)=
  3 \mathcal{L}_n (z,E_n)\, .
\end{equation}
This flux of neutrinos is directly associated with the injected neutrons and
serve as a minimal contribution from the source. In optically thicker sources
neutrons may undergo photohadronic interactions before escaping the
confinement region, as sketched in Fig.~\ref{thin} (right). This will decrease
the emissivity of neutrons compared to that of neutrinos.  Depending on the
ambient gas, $pp$ interactions may also dominate over photohadronic processes
in the source and produce additional neutrinos.

The opacity of the CMB to UHE protons propagating over cosmological distances
guarantees a cosmogenic flux of neutrinos, originated in the reaction $p +
\gamma_{\rm CMB} \rightarrow N + \pi^+ \rightarrow \mu^+\nu_\mu ...
\rightarrow\nu_\mu\bar \nu_\mu \nu_e\, e^+ ...$~\cite{Beresinsky:1969qj}.
Recently, two of us were involved in an investigation of the actual size of
the cosmogenic neutrino flux~\cite{Fodor:2003ph}.
In this work, we consider the combined flux of cosmogenic neutrinos and
neutrinos from $p \gamma$ interactions in optically thin sources, which can be
calculated by
\begin{equation}
  J_\nu(E) = \frac{1}{4\pi}\, \sum_{n,\nu} \int\limits_0^\infty \d
  E_i\int\limits_{0}^{\infty} \d r \left|\frac{\partial
  P_{\nu|i}(E;E_i,r)}{\partial E}\right|\mathcal{L}_i\, ,
\end{equation}
where the propagation functions $P_{\nu|n}$ and $P_{\nu|\nu}$ are defined
analogously to $P_{p|n}$ in the previous section.

As it was discussed in Ref.~\cite{Ahlers:2005sn} this flux of high energy
neutrinos from optically thin sources is almost in reach of the AMANDA-II
detector for neutrino energies of the order $10^7$~GeV and should soon be
observable by its successor experiment IceCube. This observation is crucial
for our assumption about the sources of CRs and the associated flux of
neutrinos from the same sources. If even cosmogenic neutrinos are not detected
in future experiments, also the assumption on the extragalactic origin of CRs
has to be reconsidered.

\section{\label{nueas}Footprints of Strongly Interacting Neutrinos}

The search for UHE cosmic neutrinos is a challenging task due to their feeble
interactions in the SM.  To overcome this problem large-scale detectors and
novel techniques are deployed and proposed. Up to now, experiments with UHE
neutrino sensitivity include AGASA, RICE, GLUE, FORTE, BAIKAL, Fly's Eye,
AMANDA,and PAO.  Further ambitious projects include ANITA and IceCube,
possibly followed by EUSO, SalSA, or OWL.
\begin{figure}[t]
\begin{center}
  \includegraphics[width=\linewidth,clip=true]{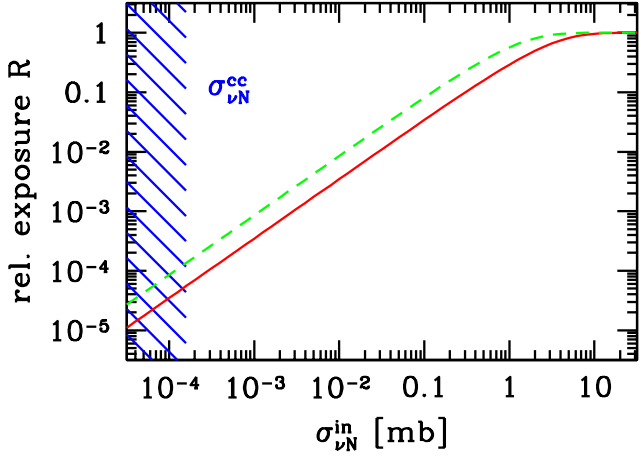}
\end{center}
\caption{\label{relexp}Relative exposure $\mathcal{R}$ (Eq.~(\ref{rel_exposure}))
  to strongly-interacting neutrinos as a function of the neutrino-nucleon
  inelastic cross section $\sigma^{\rm in}_{\nu N}$, of AGASA (solid) and
  HiRes (dashed), respectively).  The {hatched} region shows the predicted
  contribution from SM charged current
  interaction~\cite{Gandhi:1998ri,Kwiecinski:1998yf}.}
\end{figure}
However, if the neutrino-nucleon inelastic cross section $\sigma^{\rm in}_{\nu
  N} (E)$ increases more rapidly with energy than the SM predicts, UHE cosmic
neutrinos could already leave their footprints in the CR observational
data.  This happens when the neutrino interaction length $\lambda_\nu \equiv
m_p / \sigma^{\rm in}_{\nu N}$ becomes comparable with the atmospheric depth
$x_{\rm atm} (\theta)$ in the quasi-vertical direction, e.g.  $\theta \lesssim
45^\circ$ (cf. Appendix \ref{verticalshower}).
\begin{figure}[t]
\begin{center}
  \includegraphics[width=\linewidth,clip=true]{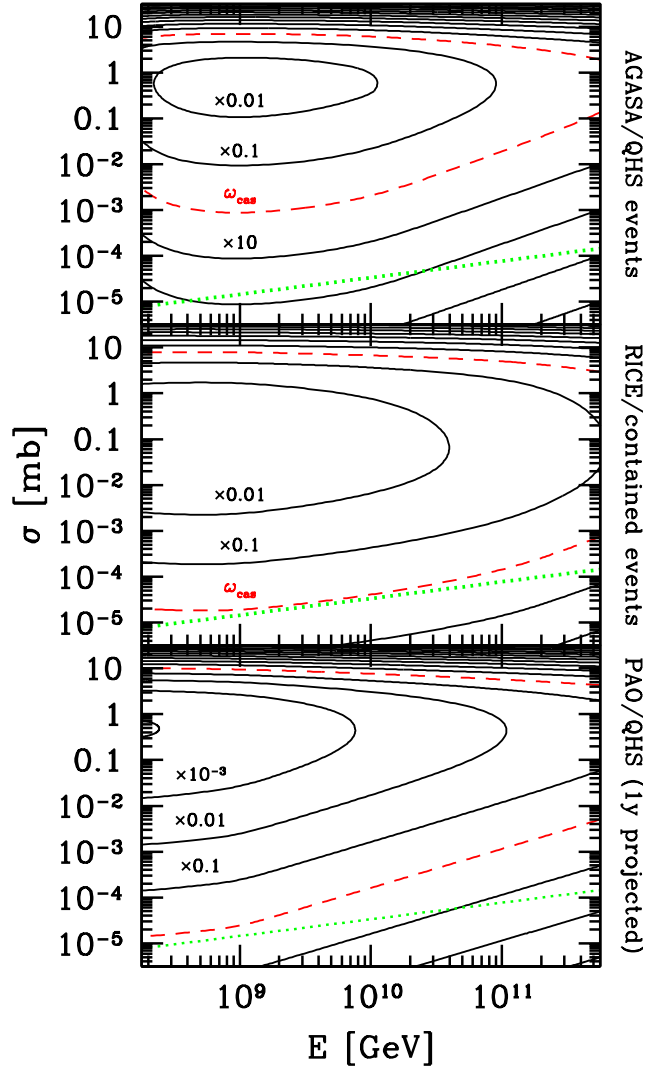}
\end{center}
\caption{\label{sensitivity}
  The sensitivity in terms of the the maximal average flux $J^\mathrm{max}$
  per bin $\log_{10}\Delta E/E = \pm0.05$, with average cross section per bin
  $\sigma$ consistent to the $95\%$ C.L. with the experimental results on QHS
  events at AGASA (top), contained events at RICE (center) and QHS events at
  PAO (bottom). The flux is shown as $E^2 J^\mathrm{max}$ relative to the
  energy density $\omega_\mathrm{cas} =
  8.5\times10^{6}\,\mathrm{eV}\,\mathrm{m}^{-2}\,\mathrm{s}^{-1}\,\mathrm{sr}^{-1}$
  ({dashed}), corresponding to the cascade limit. The contours show the
  increase of the sensitivity by one order of magnitude (comp. the Appendices
  \ref{agasa_qhs}, \ref{rice} and \ref{pao}). The {dotted} line is the
  extrapolation of SM charged and neutral current cross sections.}
\end{figure}

In this work we aim to investigate the possibility of strongly interacting
extragalactic neutrinos appearing as the highest energy CR events in
dependence of a varying $\sigma_{\nu N}^\mathrm{in} (E)$.  We use CR
data above $10^{8.6}$~GeV measured by the AGASA and HiRes collaborations, and
consider protons and neutrino primaries only.  For simplicity we assume that
the characteristics of neutrino-induced showers are indistinguishable from
those induced by protons.  In particular, we assume for both primaries {\it
  i)} a complete conversion of the incident energy into the shower, and {\it
  ii)} equal detection efficiencies at the highest energies~\cite{note1}.

Under these assumptions we define $\mathcal{R}(\sigma_{\nu N}^\mathrm{in})$ as
the relative experimental exposure to strongly interacting neutrinos compared
to the exposure $\mathcal{E}(E)$ to protons.  The number of detected events is
then given by
\begin{equation}
\label{rel_exposure}
  N_\mathrm{obs} = \int \d\,E\,\mathcal{E}(E)\left( J_p (E)+
  \mathcal{R}(\sigma^\mathrm{in}_{\nu N} (E))\, J_\nu(E)\right)\, .
\end{equation}
The relative exposure $\mathcal{R}(\sigma^\mathrm{in}_{\nu N}(E))$ is
determined by the search criterion on the zenith angle $\theta$ and the
(observed) atmospheric depth (cf.~Appendix~\ref{verticalshower} and
Fig.~\ref{slant}) adopted by each experiments. HiRes ($\theta \leq 60^\circ$)
thus has a larger relative exposure than AGASA ($\theta \leq 45^\circ$), as
Fig.~\ref{relexp} shows. For both experiments one sees that neutrinos start to
contribute significantly to quasi-vertical showers for $\sigma^{\rm in}_{\nu
  N} \gtrsim {\cal O}(1)$~mb.  On the other hand, in the intermediate range
$\sigma^{\rm SM}_{\nu N} \lesssim \sigma^{\rm in}_{\nu N} \lesssim 1~{\rm
  mb}$, neutrinos can still penetrate deeply in the quasi-horizontal
direction, or in the ice upper surface.  For a given neutrino flux, the
non-observation of such events so far can be turned into model-independent
upper bounds on the neutrino-nucleon inelastic cross section in the
intermediate range, or be used to constrain models which predict an
anomalously enhancement of $\sigma^{\rm in}_{\nu N}$ at high
energies~\cite{Anchordoqui:2004ma}.  In the following we will focus on the
search results on quasi-horizontal showers (QHS) at AGASA
\cite{Yoshida:2001pw} and contained events at RICE \cite{Kravchenko:2003tc}.
Fig.~\ref{sensitivity} shows the sensitivity of these experiments in terms of
the maximal average flux ($J^{\rm max}_\nu$) per bin and mean inelastic cross
section ($\sigma^{\rm in}_{\nu N}$) consistent with the experimental results.
For details of the calculation see Appendices \ref{agasa_qhs}, \ref{rice} and
\ref{pao}.

\section{\label{UHECS}High Energy Inelastic Neutrino-Nucleon Cross Section}

For a satisfying explanation of post-GZK events by extragalactic neutrinos,
the neutrino-nucleon cross section has to reach nucleonic values around
$E_\mathrm{GZK}$ as shown in Fig.~\ref{relexp}. In order to be consistent with
the data on QHSs at AGASA and events at RICE one expects only steeply
increasing inelastic interactions as good candidates for a combined
statistical analysis. In the following we want to support these rather
qualitative arguments by a statistical analysis of the existing data assuming
strongly interacting extragalactic neutrinos.

In frequentists statistic the level of agreement of a particular hypothesis
$\mathcal{H}$ with the experimental data can be
represented~\cite{Eidelman:2004wy} by
\begin{equation}\label{pvalue}
\mathcal{G}(\mathcal{H}) = \sum_{{\scriptscriptstyle N'|P(N')<
  P(N_\mathrm{exp})}}P(N'|\mathcal{H}) ,
\end{equation}
the integrated probability of those samples $N'$ which have a smaller
probability $P$ than the actual experimental result $N_\mathrm{exp}$. In
general, $\mathcal{H}$ is then accepted (or rejected) at a chosen significance
level $\mathcal{G}$ corresponding to a confidence level $1-\mathcal{G}$
\cite{Eidelman:2004wy}. As it is standard in statistics, we choose 90\%, 95\%
and 99\% as benchmarks for the acceptance of our model.

In our case, the probability $P$ is made up by Poisson distributions of
vertical events (AGASA, Fly's Eye Stereo and Hires-I/II; see
Appendix~\ref{verticalshower}), QHSs at AGASA (see Appendix~\ref{agasa_qhs})
and contained events at RICE (see Appendix~\ref{rice}) with an expectation
value determined by the hypothesis $\mathcal{H}$. We also account in $P$ for
the systematic error in energy calibration of about 30\% by a Gaussian
distribution for a shift of the observed spectra by a multiple of the smallest
bin-size. The expectation value for the Poisson-distributed events is
determined by the hypothesis $\mathcal{H}(\gamma,n,\sigma^\mathrm{in}_{\nu
  N})$, i.e.\ by the particular model for the inelastic neutrino-nucleon cross
section $\sigma_{\nu N}^\mathrm{in}$ together with the source luminosity
given by an injection index $\gamma$ and an evolution index $n$.

\begin{table}[t]
\caption{Best fit (81\% CL), resolution of the fit, and range of parameters.}
\begin{center}
  \begin{tabular}{c|ccccc}
\hline \hline
&$\gamma$&$n$&$\scriptstyle\log_{10}\mathcal{A}$&$\scriptstyle \log_{10}\frac{\Delta
  E}{E_{\rule[-0.2ex]{0ex}{0ex}\mathrm{th}}}$ &$\scriptstyle \log_{10}\frac{E_{\rule[-0.2ex]{0ex}{0ex}\mathrm{th}}}{1\mathrm{GeV}}$\\ 
\hline
best fit &
2.4&3.8&7.0&0.15&10.95\\
\hline \hline
grid size & 
0.01&0.05&0.16&0.044&0.1\\
min &
2.00&0.00&0.0&0.1&9.90\\
max &
2.99&4.95&7.0&1.2&11.50\\
\hline \hline
\end{tabular}
\end{center}
\label{bestfit}
\end{table}

\begin{figure}[b]
\begin{center}
  \includegraphics[width=\linewidth, clip=true]{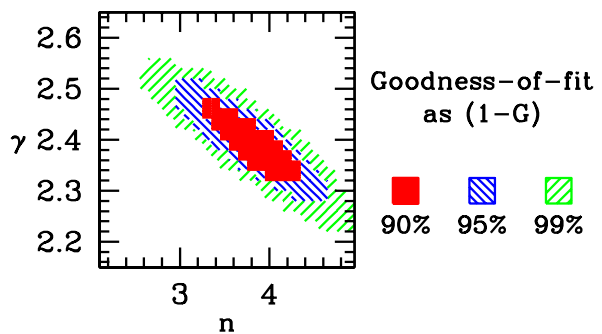}
\end{center}
\caption{\label{statinj}The $90 \%$ (shaded), $95\%$ (fine-hatched) and $99\%$
  (coarse-hatched) CLs of the injection spectrum marginalize w.r.t.\ the
  neutrino-nucleon cross section.}
\end{figure}

\begin{figure}[t]
\begin{center}
  \includegraphics[width=\linewidth, clip=true]{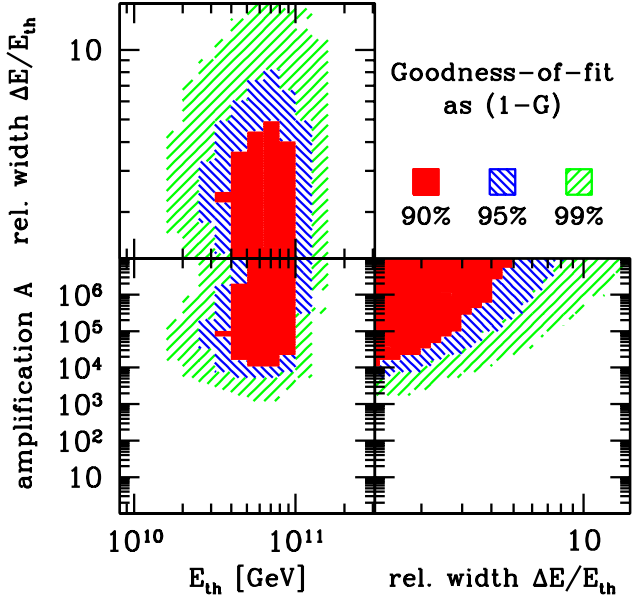}
\end{center}
\caption{\label{stat}The $90 \%$ (shaded), $95\%$ (fine-hatched) and $99\%$
  (coarse-hatched) CLs of the $\nu N$ cross section. We marginalize w.r.t.\ 
  the other parameters.}
\end{figure}

The absolute value of the predicted flux is a priori unknown due to our lack
of knowledge of the CR source luminosity. For each experiment individually we
normalize the events induced by protons and neutrinos to the data between
$10^{8.6}$~GeV and $10^{12}$~GeV. The resulting ambiguity in the normalization
of the proton and neutrino fluxes has to be removed for a prediction of
horizontal events at AGASA and contained events at RICE. In this case, we
normalize the fluxes to an average data set interpolating between AGASA, Fly's
Eye and HiRes-I/II according to the exposure of the individual bins.

\begin{figure}[b]
\begin{center}
  \includegraphics[width=\linewidth,clip=true]{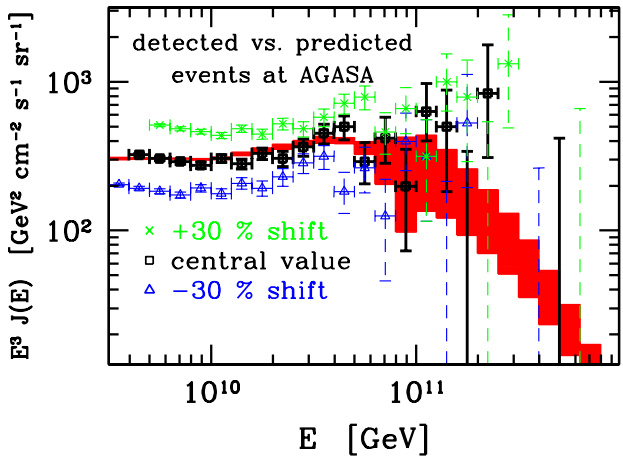}
\end{center}
\caption{\label{agasadata}The range of predicted events for AGASA induced by extragalactic
  protons and neutrinos corresponding to the 90\% CL of the goodness-of-fit
  test (shaded band) compared to the AGASA data shifted in energy by
  $\pm30\%$.}
\end{figure} 

\begin{figure}[t]
\begin{center}
  \includegraphics[width=\linewidth,clip=true]{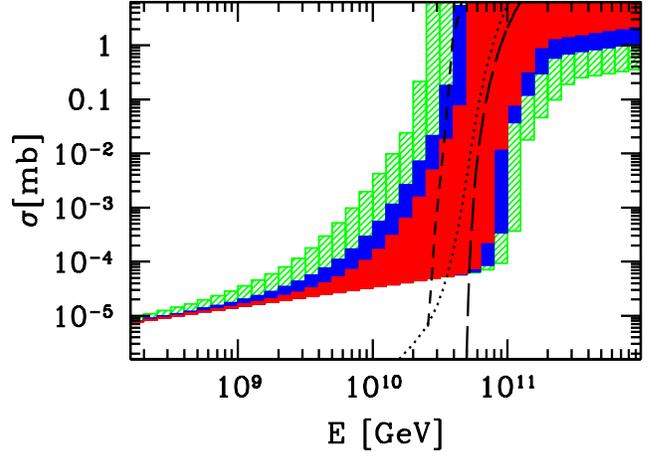}
\end{center}
\caption{\label{csbound} 
  The range of the cross section within the 99\%, 95\% and 90\% CL. The lines
  are theoretical predictions of an enhancement of the neutrino-nucleon
  cross-section by electroweak sphalerons~\cite{Han:2003ru} (short-dashed),
  $p$-branes~\cite{Anchordoqui:2002it} (long-dashed) and string
  excitations~\cite{Burgett:2004ac} (dotted).  }
\end{figure} 

Now that we have set up our statistical analysis, we need to specify the
particular neutrino-nucleon cross section for our hypothesis.  There are
various theoretical ideas for a rapid increase of the neutrino-nucleon cross
section $\sigma_{\nu N}$ referring to physics beyond the perturbative SM as
was already stated in the introduction. Based on our previous considerations,
we are interested in three characteristics of a strong neutrino--nucleon
interaction: the threshold energy $E_\mathrm{th}$ marking the changeover from
weak to strong interaction, the range $\pm \Delta E/E_\mathrm{th}$ for this
change and the final amplification $\mathcal{A}$ of the SM prediction
$\sigma_\mathrm{SM}$ of charged and neutral current interactions. We
parameterize the neutrino-nucleon cross section as
\begin{equation}\label{cs}
\log_{10}\left(\frac{\sigma_{\nu N}^\mathrm{in}}{\mathcal{A}\,\sigma_\mathrm{SM}}\right) = 
\frac{1}{2}\left[1+\tanh\left(\frac{\ln(E_\nu/E_\mathrm{th})}{\ln(\Delta E/E_\mathrm{th})}\right)\right].
\end{equation}  
The results of our goodness-of-fit test for the inelastic cross
section~(\ref{cs}) are shown in Figs.~\ref{statinj} and \ref{stat}, with the
best fit given in Table~\ref{bestfit}.  Each plot of Fig.~\ref{statinj} and
\ref{stat} shows the goodness-of-fit w.r.t.\ two parameters as confidence
levels (CLs) of $1-\mathcal{G}$. The remaining parameters of the set
$\{\gamma,n,\mathcal{A},\Delta E/E_\mathrm{th},E_\mathrm{th}\}$ are
marginalized by a $\chi^2$ minimization using simulated annealing as in
Ref.~\cite{Hannestad:2000wx}.

Comparing the boundaries of the parameters in Table~\ref{bestfit} with the
contours in Fig.~\ref{stat} it seems that the best fit is correlated to our
choice of the maximal amplification $\mathcal{A}$ and the minimal width
$\Delta E/E_\mathrm{th}$. For our statistical analysis we kept the relative
width $\Delta E/E_\mathrm{th}$ always larger than 2 in order to account for
the relative width of the bins of $\geq 10^{0.1}$. The relative exposure of
strongly interacting neutrinos shown in Fig.~\ref{relexp} indicates that the
contribution of neutrinos to the vertical spectrum saturates under an
amplification of the inelastic cross section by more than seven orders of
magnitude. This motivates us to limit the amplification below $\mathcal{A}\leq
10^7$ in our analysis.

{\it The goodness-of-fit test of the combined data requires, to the 90\% CL, a
  steep increase by an amplification factor of $\mathcal{A}>10^4$ over a tiny
  energy interval $\Delta E/E_\mathrm{th}<5$ at around $10^{11}$~GeV. In
  particular, the SM neutrino-nucleon inelastic cross section corresponding to
  $\mathcal{A}=1$ is not favored by the data.}

\begin{table}[b]
\caption{Relative contribution to the real part of the scattering amplitude at $E_\nu=100$~GeV predicted by dispersion relations.}
\begin{center}
\begin{tabular}{cccc}
\hline \hline
\,\,\,best fit\,\,\,&\,\,\,90\% CL\,\,\,&\,\,\,95\% CL\,\,\,&\,\,\,99\% CL\,\,\,\\ 
\hline
$0.091$
&$\leq0.13$
&$\leq0.15$
&$\leq0.21$\\
\hline \hline
\end{tabular}
\end{center}
\label{dr}
\end{table}

Fig.~\ref{agasadata} shows the predicted events at AGASA as a shaded band
corresponding to the 90\% CL of the fit. The observed events at AGASA are
shown together with a $30\%$ shift of the energy to higher and lower values.
The difference between the normalization of the AGASA and HiRes data can be
removed by a relative re-calibration of the energy scale by 30\%
\cite{DeMarco:2003ig,DeMarco:2005ia}. In our approach we keep the
normalization as stated from the AGASA and HiRes Collaborations and integrate
the systematic error in energy calibration into Eq.~(\ref{pvalue}).

For the $90\%$, $95\%$ and $99\%$ allowed range of the parameter shown in
Fig.~\ref{stat} we plot the range of the corresponding cross section in
Fig.~\ref{csbound}, which can be used as a {\it benchmark} test for scenarios
proposing strongly interacting neutrinos as a solution to the GZK puzzle.  As
an illustration, we have considered three models of a rapidly increasing
neutrino-nucleon cross section based on electroweak
sphalerons~\cite{Han:2003ru}, $p$-branes~\cite{Anchordoqui:2002it} and string
excitations~\cite{Burgett:2004ac}.
\begin{itemize}
\item {\it Electroweak sphalerons} : We have used the neutrino-nucleon cross
  section induced by electroweak instantons shown in Ref.~\cite{Han:2003ru},
  based on the neutrino-parton cross section from Ref.~\cite{Ringwald:2003ns},
  the latter exploiting numerical results from Ref.~\cite{Bezrukov:2003qm}. A
  direct fit of $\gamma$ and $n$ with this cross section gives $1-{\cal
    G}=98\%$, which is in very good agreement with Fig.~\ref{csbound}.
\item {\it $p$-branes} : We have calculated $\sigma(\nu N \rightarrow brane)$
  given in Ref.~\cite{Anchordoqui:2002it} as Eqs.~(9) and (10) for $m=6$ extra
  spatial dimensions, a fundamental scale of gravity $M_D = 300$~TeV and a
  ratio $L/L_* = 0.005$ of small to large compactification radii. We let all
  partons interact universally with the neutrino and use the CTEQ~\cite{CTEQ}
  Set 5D parton distribution functions (PDFs) contained in the FORTRAN library
  PDFLIB~\cite{Plothow-Besch:1992qj} Version 8.04 at the factorization scale
  $\mu=10$~TeV. Our fit of $\gamma$ and $n$ gives $1-{\cal G}=83\%$.
\item {\it String excitations} : For the neutrino-quark cross section given in
  Ref.~\cite{Burgett:2004ac} with a string scale $M_* = 70$~TeV and for the
  set of parameters $N_0 = C = 16$, characterizing the width and the absolute
  normalization, respectively (cf.\ their Fig.~2), we derived the
  neutrino-nucleon cross section with the PDFs described in the previous item.
  Our fit gives $1-{\cal G}=84\%$ for this cross section, again in very good
  agreement with Fig.~\ref{csbound}.
\end{itemize}

We should also mention at this point that a neutrino-nucleon cross section
much larger than the SM predictions at some high energy scale will also have
impact on the elastic scattering amplitude at much lower energies due to
dispersion relations~\cite{Goldberg:1998pv}. Eq.~(\ref{approx_disp}) in
Appendix~\ref{dispersion} gives the expected relative rise of the real
amplitude compared to the SM prediction for arbitrary models of the
neutrino-nucleon cross section.  We have checked that for the models displayed
in Fig.~\ref{csbound} there is no conflict with low energy data on elastic
neutrino-nucleon scattering. Table~\ref{dr} shows the maximal relative
contribution of strongly interacting neutrinos to the real part of the elastic
scattering amplitude predicted by the SM at $E_\nu=100$~GeV. For the 99\% CL
the maximal contribution is 21 \%.

\section{\label{conclusion}Conclusion and Outlook}

We have shown that current data on the highest energy cosmic rays from AGASA
and HiRes may be interpreted as the combined flux of extragalactic protons and
strongly interacting extragalactic neutrinos. For the flux of neutrinos
associated with neutrons from optically thin sources we derived requirements
on the inelastic neutrino-nucleon cross section. We found, that a sufficiently
steep increase of the cross section within one energy decade around
$E_\mathrm{GZK}$ by four orders of magnitude may serve as a model also
consistent with the search results on quasi-horizontal showers at AGASA and
contained events at RICE. Our result is summarized in Fig.~\ref{csbound} where
we show the range of the enhanced neutrino-nucleon cross section within the
90\%, 95\% and 95\% CL of the fit. We have checked that the allowed region for
the cross section is compatible with theoretical predictions, e.g.\ from
electroweak sphalerons, $p$-branes and string excitations.

Our assumption of the extragalactic origin of the ultra high energy cosmic
rays is motivated by experimental composition measurements. Standard
mechanisms for the acceleration of these particles necessitates a {\it
  minimal} flux of extragalactic neutrinos associated with the observed CRs.
If protons are produced in denser or optically thicker sources the flux of
neutrinos is expected to be increased compared to the protons and the
necessity of a strong inelastic cross section will be relaxed. This will also
be the case if we extend the cut-off of the neutron injection spectrum
$E_\mathrm{max}$ to values larger than $10^{12}$~GeV, though this seems to be
hard to achieve for astrophysical Bottom--Up sources~(see
e.g.~\cite{Torres:2004hk} for a review). These ambiguities in the flux of UHE
cosmic neutrinos will soon be clarified by future experiments such as
IceCube~\cite{Ahlers:2005sn}.

The Pierre Auger Observatory will play a crucial role on models of strongly
interacting neutrinos. Beside the spectrum of vertical showers with a much
better statistic than AGASA and HiRes, the search of quasi-horizontal showers
will soon have a stronger sensitivity to weakly interacting neutrinos as
Fig.~\ref{sensitivity} indicates. Within our approach it will be easy to
implement any future data, notably from Auger, which might finally reach large
sensitivity on strongly interacting neutrino scenarios. Here, also possible correlations with distant astrophysical sources can give a hint on neutrino primaries~\cite{Tinyakov:2001nr,Tinyakov:2001ir}. 

\section*{Acknowledgments} We would like to thank Luis Anchordoqui and Steen Hannestad for the nice
discussions, and Zoltan Fodor and Sandor Katz for comments and for the support
of our numerical analysis based on their original computer codes.


\appendix
\section{\label{verticalshower}Observation of Vertical Showers at AGASA and HiRes}

We start with the differential rate of air showers initiated at the point
$(\ell,\theta)$ by particles (cosmic ray protons, cosmic neutrinos etc.)
incoming with energy $E$ and of flux $J(E)$. Here, $\ell$ is the distance of
this point to the detector center measured along the shower axis, and $\theta$
is the angle to the zenith at the point the shower axis hits the Earth's
surface. We will focus in the following on showers with axis going through the
detector. For detectors using the Fly's Eye technique this slightly
underestimates the rate of induced event. However, for the relative exposure
${\cal R}$ of strongly interacting neutrinos this should effect the results
only for small cross sections, and then by a negligible amount.

The number of \emph{induced} showers due to the inelastic interaction, the
cross section for which is $\sigma^{\rm in}$, per unit of length $\ell$ along
the shower axis, time $t$, area $A_\perp$ perpendicular to the shower axis,
shower energy $E_\mathrm{sh}$ and solid angle $\Omega$ (with $d\Omega =
\sin\theta\, d \theta\, d\phi$)) is
\begin{multline}
  \frac{\d}{\d \ell} \left(\frac{\d^4 N_\mathrm{ind}}{\d t\,\d
      A_\perp\,\d\Omega\,\d E_\mathrm{sh}}\right)  \\
  = \frac{\rho_\mathrm{air}\left[h(\ell, \theta) \right]}{m_p}\,
  \sigma^\mathrm{in}(E)\, J(E)\, e^{-\frac{\sigma^\mathrm{in}(E)\,
      x(\ell,\theta)}{m_p}}\, ,
\end{multline}
where $m_p$ is the proton mass, and $\rho_{\rm air}$ is the air density at the
altitude $h$.  The energy deposited in a visible shower is related to the
incident particle energy by the inelasticity parameter, i.e. $E_{\rm sh} = y
E$.  The distribution of $y$ is dependent on the scattering process.

\begin{figure}[t]
\begin{center}
  \includegraphics[width=\linewidth,clip=true]{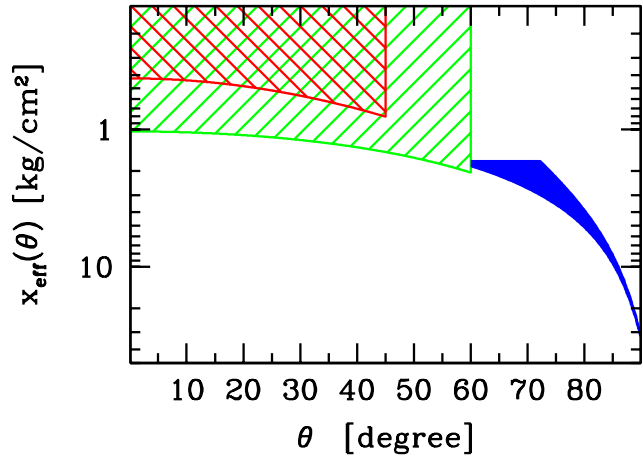}
\end{center}
\caption{\label{slant}The range of atmospheric depths $x_-<x_\mathrm{eff}<x_+$ 
  used for the calculation of Eq.~(\ref{observed}) for quasi-vertical showers
  at AGASA ({left-hatched}) and HiRes ({right-hatched}) and quasi-horizontal
  showers at AGASA ({shaded}).  For the quasi-vertical showers we use $x_-
  \approx 0\,\mathrm{kg}\,\mathrm{cm}^{-2}$.}
\end{figure}

The number of showers \emph{observed} is determined by the trigger efficiency
$\mathcal{P}(E,\theta)$ of the detector, and the range of atmospheric depths
within which showers induced are visible to the detector.  The latter, after
carrying out the integral $\rho_{\rm air} (\ell, \theta)\, d\ell \equiv - dx$,
translates into a minimal and maximal atmospheric depth $x_-(\theta)$ and
$x_+(\theta)$, as shown in Fig.~\ref{slant}.  After integration along the line
of sight of the observer, one has
\begin{multline}\label{observed}
  \frac{\d^4 N_\mathrm{obs}}{\d t\,\d A_\perp\,\d\Omega\,\d E_\mathrm{sh}}
  \\
  = \mathcal{P}(E_\mathrm{sh},\theta)\,J(E)
  \left(e^{-\frac{\sigma^\mathrm{in}(E)\,x_-(\theta)}{m_p}}-
    e^{-\frac{\sigma^\mathrm{in}(E)\,x_+(\theta)}{m_p}} \right)\, .
\end{multline}
From this we can read off the experimental \emph{exposure}
$\mathcal{E}(E)\,\,[\mathrm{s}\,\mathrm{m}^2\,\mathrm{sr}]$ defined as:
\begin{equation}
\label{exposure}
   N_\mathrm{obs} = \int \d\,E\,\mathcal{E}(E)J(E)\, .
\end{equation} 
The AGASA and HiRes spectra shown in Fig.~\ref{data} have been observed in the
quasi-vertical direction with $\theta<45^\circ$ and $\theta<60^\circ$,
respectively. In this angular range both detectors are sensitive to
interactions in the outermost atmospheric layers and $x_-(\theta)$ decreases
to zero. The maximal atmospheric depth $x_+(\theta)$ depend on the height of
the detector sites and the line of sight observed in the atmosphere. Also, for
the case of AGASA we reduce the total atmospheric depth $x(\theta)$ by
$500\,\mathrm{g}\,\mathrm{cm}^{-2}$ in order to exclude those interactions too
close to the detector site in order to be triggered \cite{Morris:1993wg}.
Note that on average a $10^{19}$~eV shower develops to its maximum after
traversing $800\,\mathrm{g}\,\mathrm{cm}^{-2}$ in the atmosphere.

\section{\label{agasa_qhs}Quasi-horizontal Showers at AGASA}

The AGASA Collaboration has observed one quasi-horizontal ($\theta>60^\circ$)
shower (QHS) during an operation time of 1710.5 days with an expected
background of $1.72$\ou{+0.14+0.65}{-0.07-0.41} (MC statistics and
systematics) \cite{Yoshida:2001pw}. The search criteria set the following
constraints on the shower maximum $x_\mathrm{max}$:
$x(\theta)-x_\mathrm{max}(\theta) < 500\,\mathrm{g}\,\mathrm{cm}^{-2}$ and
$x_\mathrm{max} \geq 2500\,\mathrm{g}\,\mathrm{cm}^{-2}$. On average the
shower develops its maximum after traversing
$800\,\mathrm{g}\,\mathrm{cm}^{-2}$ in the atmosphere. Hence the observed
atmospheric depth shown in Fig.~\ref{slant} varies between $x_-(\theta) =
\max(1700\,\mathrm{g}\,\mathrm{cm}^{-2},x(\theta)-1300\,\mathrm{g}\,\mathrm{cm}^{-2})$
and $x_+(\theta) = x(\theta)$. The number of observed QHS events can then be
calculated by Eq.~(\ref{observed}). As the effective detector area for
hadronic showers we took $A=56.1$~km$^2$. The horizontal detection efficiency
$\mathcal{P}_\mathrm{hor}(E)$ for QHSs is reported to be 100\% above
$10^{10}$~GeV and approximately zero below $10^{8}$~GeV. In between we use a
power-law approximation $\propto(\log_{10}(E/\mathrm{GeV})-8)^n$, which is
fitted \cite{Tu:phd} to reproduce the upper bound of 3.52 events (95\% CL) for
the observation of one QHS from charged current interactions reported by the
AGASA Collaboration.  The upper plot in Fig.~\ref{sensitivity} shows the
sensitivity of AGASA for QHSs in terms of the maximal neutrino flux
$E^2_\nu\,J^\mathrm{max}_\nu(E_\nu)$ per flavor and per energy range
$\log_{10}\triangle E/E = \pm0.05$ consistent to the 95\% CL with the
observation.

\section{\label{rice}Contained Events at RICE}

The Radio Ice Cherenkov Experiment (RICE) at the South Pole has searched for
electro-magnetic and hadronic showers based on the principle of ``radio
coherence''. During an observation time of $3500$ hours no candidate of an
neutrino-induced event has been observed \cite{Kravchenko:2003tc}. The
expected number of events can be approximated as:
\begin{equation}\label{coneven}
\frac{\d^3\,N}{\d t\,\d\Omega\,\d E}
=\frac{\rho_\mathrm{ice}}{m_p}\int\limits_{V_\mathrm{eff}(E)}\!\d\vec{r}\,
J_\nu(E)\, \sigma^\mathrm{in}_{\nu N}(E)\, e^{-\frac{\sigma^\mathrm{in}_{\nu
      N}\,x(\vec{r},\theta)}{m_p}}
\end{equation}

The effective detection volume $V_\mathrm{eff}(E)$ has been determined by MC
simulations in \cite{Kravchenko:2002mm}. We approximate the ice target as a
cylinder with $V_\mathrm{eff}(E)=h\pi r^2(E)$ and a fixed height
$h=1\,\mathrm{km}$.  From this we can approximate the distance
$d(\vec{r},\theta)$ a quasi-horizontal neutrino has to traverse in ice before
it interacts with a nucleon inside the detector. Hence, the depth $x(\theta)$
is composed of the atmospheric depth $x_\mathrm{atm}(\theta)$ and the depth in
ice $d(\vec{r},\theta)\rho_\mathrm{ice}$. The sensitivity of RICE to the
neutrino flux is shown as the center plot in Fig.~\ref{sensitivity}.

\section{\label{pao}Quasi-Horizontal Showers at PAO}

The Pierre Auger Observatory (PAO), which is actually comprised of two
sub-observatories, is the next large-scale neutrino detector in operation.
The Southern site is currently operational and growing to its final size of
$\simeq 3000~{\rm km}^2$.

The rate of neutrino-induced events at the ground arrays of PAO can be
calculated using Eq.~(\ref{observed}).  The effective aperture has been
parametrized in Ref.~\cite{Anchordoqui:2004ma} through a comparison with the
geometric acceptance published in Ref.~\cite{Capelle:1998zz}.  In short, to
estimate the sensitivity for PAO, the following selection criteria were
adopted: {\it i)} $75^\circ \leq \theta \leq 90^\circ$ for the zenith angle,
{\it ii)} $X_{\rm max} \geq 2500~{\rm g/cm^2}$ for the shower maximum, which
corresponds to requiring $x_-(\theta) = 1700 \mathrm{g}/\mathrm{cm^2}$ in this
work.  The altitude of the PAO Southern site (1200~m above sea level) was also
taken into account in $x_+(\theta) = x(\theta)$.  For hadronic showers with
axis falling in the array, the effective area can then be parametrized as
$A_\perp (\theta, E)\, P(E)$, with $A_\perp (\theta,E) \approx \cos \theta\,
\cdot 1.475~\mathrm{km^2}\, (E/\mathrm{eV})^{0.151}$, and $P(E)=1$ for $E \geq
10^{8.6}$~GeV, while $P(E) = 0.654~\log_{10}(E/\mathrm{eV})-10.9$ below this
energy.  The effective aperture for all showers (i.e. including showers with
axis not going through the array) is roughly 1.8 to 2.5 times larger, as shown
in Ref.~\cite{Capelle:1998zz}.

A first model-independent investigation of the sensitivity of PAO to anomalous
neutrino interactions was performed in Ref.~\cite{Anchordoqui:2004ma}.
Assuming one year of operation with no event observed above the expected
hadronic SM background (95\% CL corresponding to 3.09 events), we estimate the
prospects for PAO to strengthen the existing constraints imposed by AGASA and
RICE.  The 1-year projected sensitivity is shown in Fig.~\ref{sensitivity}
(lowest panel).

\section{\label{dispersion}Dispersion Relations}

Regardless of a particular model, high energy cross section have to fulfill
criteria relying on the analyticity and unitarity of the S-matrix. As was
emphasized in \cite{Goldberg:1998pv} the total cross sections at high energies
are constrained by low energy elastic amplitudes due to dispersion relations.
One can relate the elastic scattering amplitude~\cite{note2} $\Re A$ to the
principal value of an integral involving the total (anti-) neutrino-nucleon
cross section $\sigma^\mathrm{tot} = \sigma^\mathrm{SM} +
\sigma^\mathrm{new}$:
\begin{multline}\label{disp}
  \Re A(E_\nu) - \Re A(0) =\\
  \frac{E_\nu}{4\pi}\mathcal{P}\int_0^\infty \d E' \left(\frac{\sigma_{\nu
        N}^\mathrm{tot}(2m_pE')}{E'(E'-E_\nu)} +
    \frac{\sigma_{\bar{\nu}
        N}^\mathrm{tot}(2m_pE')}{E'(E'+E_\nu)}\right)
\end{multline}
We assume that our high energy cross section obeys the Pomerantschuk theorem,
i.e. $\sigma^\mathrm{new}_{\nu
  \mathrm{N}}-\sigma^\mathrm{new}_{\bar{\nu}\mathrm{N}} \rightarrow 0$ for
$E_\nu \rightarrow \infty$. The elastic amplitude at $E_\nu \approx 0$ is
dominated by Z-boson exchange of the order of $G_F/2\sqrt{2}$. For $E_\nu \ll
E_- \ll E_\mathrm{th}-\Delta E $ we can use Eq.~(\ref{disp}) to estimate the
relative contribution of new physics at low energies as:
\begin{equation}\label{approx_disp}
\frac{\Re A_\mathrm{new}(E_\nu)}{\Re A_\mathrm{SM}(E_\nu)} \approx \frac{\sqrt{2}E_\nu}{0.637\pi G_\mathrm{F}}\int\limits^\infty_{E_-}\d\, E' \frac{\sigma^\mathrm{SM}}{E'}\frac{\d}{\d\,E'}\left(\frac{\sigma^\mathrm{tot}}{\sigma^\mathrm{SM}}\right)
\end{equation}



\begin{thebibliography}{99}
\bibitem{note2} We do not consider helicity of the nucleos in the following.
\end{thebibliography}
\end{document}